\documentclass[prl,superscriptaddress,unsortedaddress,twocolumn,%
  showpacs,preprintnumbers,amsmath,amssymb]{revtex4-1}
\usepackage{graphicx}
\usepackage[breaklinks,colorlinks=true]{hyperref}

\usepackage{color}
\definecolor{darkgreen}{rgb}{0.0,0.6,0.0}
\definecolor{violet}{rgb}{0.6,0.4,0.8}

\def\Msun{M_\odot}

\begin{document}
\title{Methodology study of machine learning for the neutron star equation of state}
\author{Yuki Fujimoto}
\author{Kenji Fukushima}
\author{Koichi Murase}
\affiliation{Department of Physics, The University of Tokyo, %
  7-3-1 Hongo, Bunkyo-ku, Tokyo 113-0033, Japan}

\begin{abstract}
  We discuss a methodology of machine learning to deduce the
  neutron star equation of state from a set of mass-radius observational
  data.  We propose an efficient procedure to deal with a mapping from
  finite data points with observational errors onto an equation of
  state.  We generate training data and optimize the neural network.
  Using independent validation data (mock observational data) we
  confirm that the equation of state is correctly reconstructed with
  precision surpassing observational errors.  We finally discuss the
  relation between our method and Bayesian analysis with an
  emphasis put on generality of our method for underdetermined
  problems.
\end{abstract}
\maketitle


The equation of state (EoS) of dense nuclear and quark matter should
be derived from quantum chromodynamics (QCD), but the
sign problem in dense QCD prevents us from the first
principles calculation~\cite{Aarts:2015tyj}.  It is unlikely that the
conventional nuclear EoS in terms of nucleons keeps
validity in deep cores of the neutron star (see
Ref.~\cite{Baym:2017whm} for a review).  If we use a typical
nuclear EoS to realize a two solar mass neutron
star~\cite{Demorest:2010bx,*Fonseca:2016tux,*Antoniadis:2013pzd}, the
central density could exceed several times $\rho_0$, where
$\rho_0$ represents the nuclear mass density at the saturation
point, i.e.\ $\rho_0\simeq \text{(nucleon mass)} \times
0.16\,\text{[nucleon/fm$^3$]}\simeq 2.7\times 10^{14}\,\text{[g/cm$^3$]}$.
Novel phases of matter are expected at such high density, but there is no
established description for a transition between various
matter.  High temperature QCD phases (see
Ref.~\cite{Fukushima:2010bq,*Schmidt:2017bjt} for recent reviews) have
inspired a continuous crossover scenario from nuclear to quark degrees
of freedom~\cite{Masuda:2012kf,*Alvarez-Castillo:2013spa,*Fukushima:2015bda},
which is called quark-hadron
continuity~\cite{Schafer:1998ef,*Alford:1999pa,*Fukushima:2004bj,*Hatsuda:2006ps}.

Our knowledge is limited and we need scenario independent
approaches to neutron star studies.  To this end experimental
information would be useful to constrain possible EoS
candidates.  We have such valuable experimental data of the
neutron star mass $M$ and radius $R$, and the $M$-$R$ points from
neutron star observations would ideally shape a curve called
the $M$-$R$ relation~\cite{Ozel:2016oaf}.

The one-to-one correspondence between the $M$-$R$ relation and the EoS
is, \textit{formally}, mediated by the Tolman-Oppenheimer-Volkoff (TOV)
equation~\cite{Tolman:1939jz,*Oppenheimer:1939ne} coupled with
$dm/dr=4\pi\rho r^2$ where $r$ is the radial distance, $\rho$
is the mass density, and $m$ is the mass within the radius-$r$ sphere.
Specifically, the EoS refers to $\rho=\rho(p)$ with $p$ being the
pressure (where neglecting rotation and magnetic effects are
assumed; see Ref.~\cite{Sotani:2014rva} for a modified EoS).  The TOV
equation is a differential equation for $p(r)$ and $\rho(r)$.
The radius $R$ is fixed by $p(r=R)=0$, and the mass is
given by $M=m(R)$.  It is possible to solve the TOV
equation from the $M$-$R$ relation to the EoS (up to some critical
density) as discussed in Ref.~\cite{Lindblom:1992}.

However, \textit{practically}, we do not know the $M$-$R$ relation
with arbitrary precision from neutron star observations.  To
complicate matters, a third family scenario may be
realized~\cite{Schertler:2000xq,Alford:2017qgh}.  Thus, instead of
revealing the unique EoS from the $M$-$R$ curve, we construct the most
likely EoS from discrete $M$-$R$ points.  We should develop a
robust approach to deal with observational $M$-$R$ points that deviate
from the genuine $M$-$R$ relation with errors.

One strategy is as follows.  First, we setup an EoS with
several parameters (such as parametrized spectral function of
relativistic enthalpy~\cite{Lindblom:2010bb,*Lindblom:2013kra},
piecewise polytropic parametrization~\cite{Read:2008iy}, etc).  Then,
we proceed to determine parameters by making the outputs closest to
the observational data.  This approach works for the current problem
with only discrete observational points, but it is nontrivial how
to estimate the parametrization dependence systematically.

It would be desirable to establish some alternative method in a
systematic way.  Along these lines, recently, a method based on the
Bayesian analysis has attracted theoretical
interest~\cite{Ozel:2010fw,*Raithel:2017ity,Steiner:2010fz,Alvarez-Castillo:2016oln}.
In Bayesian analysis a certain prior distribution of EoS is
prepared, and Bayesian updating for the EoS distribution is made
by the $M$-$R$ observations.  The EoS parametrization dependence is
incorporated in the prior dependence, and can be quantified by
comparing different priors.  In principle, if the number of the
$M$-$R$ data points is sufficiently large, the prior distribution
dependence can be arbitrarily suppressed.

The purpose of this work is to address another method, which is
complementary to Bayesian analysis and is straightforwardly
implemented numerically.  We will introduce a new
principle to infer the neutron star EoS utilizing deep
(i.e.\ many-layered) neural network of machine learning, which
has been successfully applied to QCD and nuclear
physics~\cite{Pang:2016vdc,Niu:2018csp}.
Throughout this paper, we use the natural unit; $c=G=1$.
\vspace{0.3em}


Here, we make a brief overview on machine learning and deep neural
network.  This method provides a handy and powerful way to find
an optimized mapping expressed in the ``neural network'' model.  For
the ``supervised'' learning, we first prepare ``training data'', that
is, data sets of input and output, and then optimize the
parameter set of the mapping from input to output.
Once the optimization is sufficiently achieved or the training is
complete, the neural network model can conversely make an educated
guess about the most likely output corresponding to a given input.
The advantage of machine learning, as compared to ordinary fitting
procedures, is that we need not rely on preknowledge about fitting
functions because the multi layer structures are capable of capturing
any functions.

\begin{figure}
  \centering
  \includegraphics[width=0.9\columnwidth]{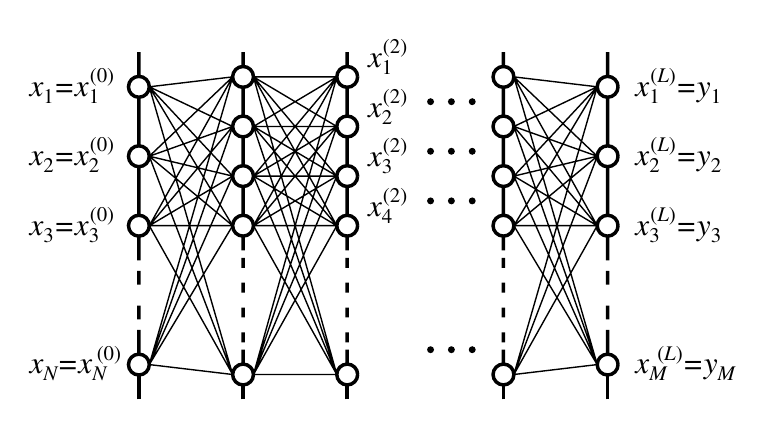}
  \caption{Feedforward neural network.}
  \label{fig:ff-neural-network}
\end{figure}

The model function of feedforward neural network can be
expressed as follows:
\begin{equation}
  y_i = f_i(\{x_j\}|\{W_{jk}^{(1)},a_j^{(1)},\dots,W_{jk}^{(L)},a_j^{(L)}\})\,,
\end{equation}
where $\{x_i\}$ and $\{y_i\}$ are input and output data, respectively.
We setup $L+1$ layers (including the input and the output layers).
Fitting parameters, $\{W_{ij}^{(k)},a_i^{(k)}\}$, on the $k$-th layer,
denote the weights between nodes in two adjacent layers and the
activation offset at each node [see Eq.~\eqref{eq:sigma}].
For the zeroth layer, the input is set as
$x_i^{(0)} = x_i$ $(1\le i\le N_1)$ with $N_1$ being the size of
input $\{x_i\}$.  For the subsequent layers, the transformations
are iteratively applied as
\begin{equation}
  x_i^{(k+1)} = \sigma^{(k+1)} \Biggl( \sum_{i=1}^{N_k}W_{ij}^{(k+1)}
    x_j^{(k)} + a_i^{(k+1)} \Biggr)\,,
  \label{eq:sigma}
\end{equation}
which defines $f_i$, where $1\le i\le N_{k+1}$ with $N_{k+1}$ being
the node numbers.  The final output from the $L$-th layer
is $y_i = x_i^{(L)}$ ($1 \le i \le N_L$) with $N_L$ being the size of
output $\{y_i\}$.  Here, $\sigma^{(k)}(x)$'s are called
``activation functions'' and the typical choices include the sigmoid
function $\sigma(x) = 1/(e^x+1)$, the ReLU $\sigma(x) = \max\{0, x\}$,
hyperbolic tangent $\sigma(x)=\tanh(x)$, etc.  The general design
structure is schematically depicted in
Fig.~\ref{fig:ff-neural-network}, in which the calculation proceeds
from the left with input $\{x_i\}$ to the right with output $\{y_i\}$.

For the actual optimization procedure we choose a
``loss function'' to be minimized;  if the loss function is
the mean square deviation, the learning amounts to the standard least
square method with $f_i$ expressed by neural network.
\vspace{0.3em}


For better learning, the training data quality is important.
For the training purpose we should
not bias data from physical reasonability, and intuitively
unnatural data should be also included to raise more adaptive neural network.
Now, let us explain how we have prepared training data which consist
of randomly generated EoS and corresponding observational points,
($M_i$, $R_i$).

First, we elucidate our scheme for the EoS generation (see
Ref.~\cite{Raithel:2016bux} for details).  Up to the density $\rho_0$,
we use a conventional nuclear EoS
(i.e.\ SLy~\cite{Douchin:2001sv} in this study), and a range
$[\rho_0,\, 8\rho_0]$ is equally partitioned in logarithmic scale into
five segments.  We randomly assign the average sound velocity
$dp/d\rho = c_s^2$ to five segments according to the uniform
distribution within $0.02<c_s^2<0.98$ where a small margin by $0.02$
is a regulator to avoid singular behavior of the TOV equation.  From
these sound velocities we determine the pressure values at segment
boundaries.  We interpolate the EoS inside of each segment
assuming polytrope $p \propto \rho^{\Gamma}$.  We note that we allow for
small $c_s^2$ corresponding to a (nearly) first-order phase
transition.  We generated 2000 EoSs in this way.

Next, we solve the TOV
equation~\cite{Tolman:1939jz,*Oppenheimer:1939ne} using the generated
$p(\rho)$ from $m=r=0$ and the enthalpy
density $h=h_c$ (where $h_c$ is a free parameter corresponding to a
choice of the central core density) until $h$ hits zero (see
Ref.~\cite{Lindblom:1992} for the formulation using $h$).  Then, we
identify $M=m(h=0)$ and $R=r(h=0)$, so that ($M$, $R$) with
various $h_c$ gives the $M$-$R$ curve.  For each randomly generated
EoS we get the $M$-$R$ curve and identify the maximum mass
$M_{\rm max}$.  If $M_{\rm max}$ does not reach the observed mass
[i.e.\ $1.97\Msun$ from the
  lower bound of $(2.01\pm 0.04)\Msun$~\cite{Antoniadis:2013pzd} where
  $\Msun$ denotes the solar mass], such EoSs are rejected from the
ensemble.  In this work 52 out of 2000 EoSs are rejected (1948
remaining).

Then, for each EoS and corresponding $M$-$R$ relation, we sample 15
observational $(M_i, R_i)$.  Here, this choice of 15 is
simply for the demonstration purpose, so it should be adjusted
according to the number of available neutron star observations (which
is so far 18 and increasing in the future~\cite{Ozel:2016oaf}).  For
better training quality, we should make unbiased sampling of 15 data
points, and we assume a uniform distribution of $M$ over
$[\Msun,\, M_{\rm max}]$.  If there are multiple values of
$R$ corresponding to one $M$, we always take larger $R$
discarding unstable branches.  In this way, we select 15 points of
$(M^{(0)}_i, R^{(0)}_i)$ on the $M$-$R$ relation.  We also train
neural network to learn that real observational data contain errors,
$\Delta M$ and $\Delta R$, which makes data points departed away from
the genuine $M$-$R$ relation.  We randomly generate $\Delta M_i$
and $\Delta R_i$ according to the normal distribution with variances,
$0.1\Msun$ and $0.5\,\text{km}$ for the mass and the radius,
respectively (as chosen in accord with next generation
measurements~\cite{Ozel:2010fw,*Raithel:2017ity}).  The variances
should also be adjusted according to the real error estimate
from observations.  Now we obtain the training data set,
$(M_i=M^{(0)}_i+\Delta M_i, R_i=R^{(0)}_i+\Delta R_i)$.  We call this
pair of $M_i$ and $R_i$ an ``observation''.  We repeat
this procedure to make 100 observations (denoted by $n_s$ later,
and the choice of $n_s$ is arbitrary if large enough for learning)
for each EoS, and finally,
we have prepared
$(1948\text{ EoSs})\times (n_s\text{ observations})=194800$ training
data in this work.
\vspace{0.3em}

\begin{table}
  \centering
  \begin{tabular}{|c||c|c|} \hline
  Layer index & Nodes & Activation \\ \hline\hline
  0 & 30 & N/A \\ \hline
  1 & 60 & ReLU \\ \hline
  2 & 40 & ReLU \\ \hline
  3 & 40 & ReLU \\ \hline
  4 & 5  & $\tanh$ \\ \hline
  \end{tabular}
  \caption{Our neural network design in this work.  In the zeroth layer
    30 nodes correspond to input 15 points of the mass and the radius.
    In the last layer 5 nodes correspond to 5 output parameters of
    the EoS.}
  \label{tab:neural-network}
\end{table}


For numerics we make use of a Python library,
Keras~\cite{software:Keras} with TensorFlow~\cite{arXiv:1605.08695} as
a backend.  The design of our neural network is summarized in
Tab.~\ref{tab:neural-network}.  Our purpose is to construct neural
network that can give us one EoS in the output side in response to one
observation, ($M_i$, $R_i$) ($i=1,\dots 15$) in the
input side.  Thus, in the zeroth layer 30 nodes should match 15 $M$-$R$
points (30 input data).  For the practical reason we
sort 30 data points by their masses in ascending order.  The output
nodes for the prediction target in the last layer correspond to 5
(sound velocity) parameters characterizing an EoS{}.  We find that the
learning proceeds faster if data are normalized appropriately;  we
use $M_i/M_{\rm norm}$ and $R_i/R_{\rm norm}$ with
$M_{\rm norm}=3\Msun$ and $R_{\rm norm}=20\,\text{km}$.

We choose the activation function at the output layer
as $\sigma^{(4)}(x) = \tanh(x)$ since the speed of sound is
automatically bounded in $[0, 1]$.  For other
layers we choose the ReLU, i.e.\ $\sigma^{(k)}(x) = \max\{0, x\}$
($k=1,2,3$), which is known to evade the vanishing gradient problem.
We specify the loss function as \texttt{msle}, that is, the mean
square $\log$ of prediction errors and choose the fitting method
as Adam~\cite{DBLP:journals/corr/KingmaB14} with the batch size 100.
To capture the essence of the problem, the complexity of layers and
nodes should be sufficiently large.  Simultaneously, to avoid the
overfitting problem, and to train neural network within a reasonable
time, the number of layers and nodes
should not be too large.  We found good performance with the node
numbers greater than the input node number on the first layer.
\vspace{0.3em}


The neural network is optimized to fit the training data, but it must have a
predictive power for independent data.  To test it, we need
``validation data'' which can be regarded as mock data for the neutron
star observation.  We generate 200 EoSs, among which 196 EoSs pass the
massive neutron star condition.  We sample just one observation for
each EoS, unlike 100 observations for training data, to mimic real
observational situations.

\begin{figure}
  \centering
  \includegraphics[width=\columnwidth]{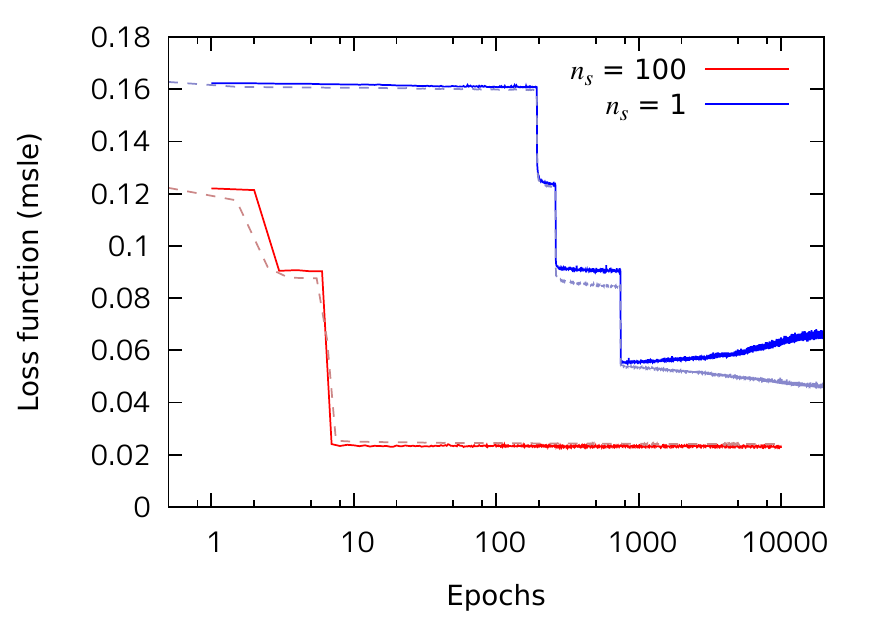}
  \caption{Loss function estimated for the training data (dashed
    lines) and the validation data (solid lines) as functions of the
    epoch.  The observation number is denoted by $n_s$.}
  \label{fig:training-history}
\end{figure}

Figure~\ref{fig:training-history} shows typical behavior of the loss
function for the training data (dashed lines) and the validation data
(solid lines) as a function of training time in units of epoch which
represents a single scan of the entire training data.  The red solid
and dashed lines show the results with $n_s=100$, i.e.\ 194800 data
set, where $n_s$ is the observation number per EoS{}.  The dashed line
is the loss function for the training data minimized through learning,
and the solid line is the loss function for the validation data
showing the performance of neural network.  We monitor the
whole history of these quantities over epochs, which is useful to
judge when the training is optimally stopped before overfitting.  We
see that the training is completed within 10 epochs for this example
in Fig.~\ref{fig:training-history}.  For the test purpose to see the
efficiency improved by $n_s$, we also show results with $n_s=1$ by the
blue solid and dashed lines in Fig.~\ref{fig:training-history}.  The
faster learning with $n_s=100$ than $n_s=1$ can be explained by data
set sizes (194800 for $n_s=100$ and 1948 for $n_s=1$).  It is
important to emphasize that introducing large $n_s$ in our proposal
can reduce the computational cost needed to increase the
data set size.  Interestingly, moreover, the validation loss function
for $n_s=1$ shows overfitting; in general, the loss function for the
training data monotonically decreases.  For the validation data,
however, it may not necessarily decrease and increasing behavior is
seen for $n_s=1$ for epochs $\gtrsim 1000$.  This significant
separation of training and validation loss functions signals
overfitting and then the predicted output could largely deviate from
the true answer.  We learn from Fig.~\ref{fig:training-history} that
the overfitting problem is also cured by $n_s \gg 1$.

\begin{figure}
  \centering
  \includegraphics[width=\columnwidth]{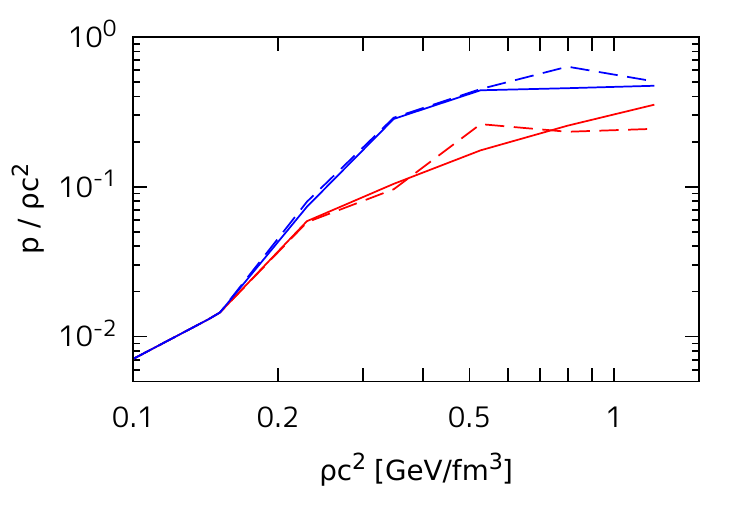}
  \caption{Two examples of the randomly generated EoSs (dashed lines)
    and the machine learning outputs (solid lines) reconstructed from
    one observation of 15 $M$-$R$ points [see
    Fig.~\ref{fig:inferred-mr} for actual ($M_i$, $R_i$)].}
  \label{fig:inferred-eos}
\end{figure}

\begin{figure}
  \centering
  \includegraphics[width=0.9\columnwidth]{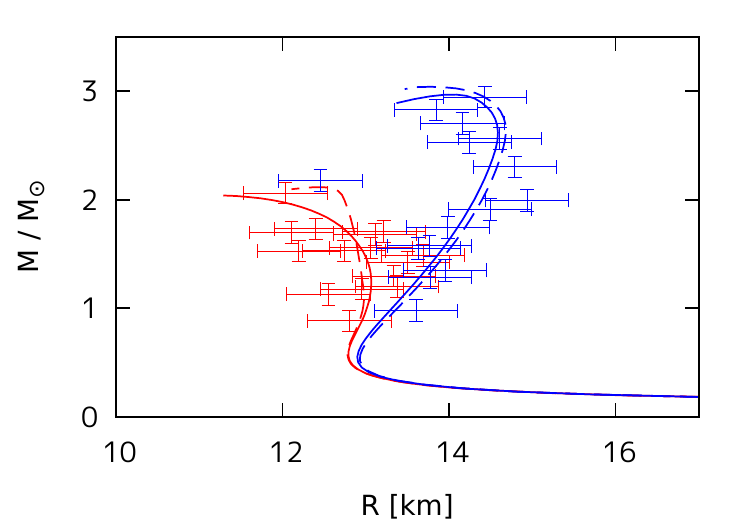}
  \caption{Randomly sampled 15 data points and the $M$-$R$ relations
    with the reconstructed EoS (solid lines) and the original EoS
    (dashed lines).  The red and blue colors correspond to two EoSs
    shown with the same color in Fig.~\ref{fig:inferred-eos}.}
  \label{fig:inferred-mr}
\end{figure}

Once the loss function converges, we can use the trained neural
network to infer an EoS from an observation of 15 $M$-$R$
points.  We picked two examples for Fig.~\ref{fig:inferred-eos}.
Later, we will quantify the overall performance and for the moment we
shall discuss these examples.  In Fig.~\ref{fig:inferred-eos} the
dashed lines represent randomly generated EoSs.  We see that two EoSs
are identical in the low density region because SLy is
employed at $\rho\leq\rho_0$.  We sampled 15 points
as shown in Fig.~\ref{fig:inferred-mr}, which mimic an observation
with error deviations from the genuine $M$-$R$
relation (which is shown by the dashed lines).  Thus, each set of 15
points is considered as mock data of the neutron star
observation.  Since the neural network learns through the training
data that the observation contains errors, the most likely EoS is
reconstructed from one observation of 15 points with errors.
The reconstructed EoSs are depicted by solid lines in
Fig.~\ref{fig:inferred-eos}.  We can see that the reconstructed EoSs
agree quite well with the original EoSs for these examples.  It
would also be interesting to make a comparison of the $M$-$R$
relations corresponding to the original and reconstructed EoSs.  The
solid and dashed lines in Fig.~\ref{fig:inferred-mr} represent the
$M$-$R$ relations calculated with the original and reconstructed
EoSs, respectively.  Since the EoSs look consistent
in Fig.~\ref{fig:inferred-eos}, the original and reconstructed
$M$-$R$ relations are close to each other.

\begin{table}
  \begin{tabular}{c||c|c|c|c|c|c|c} \hline
    Mass ($\Msun$) & 0.6 & 0.8 & 1.0 & 1.2 & 1.4 & 1.6 & 1.8 \\ \hline
    RMS (km) & 0.16 & 0.12 & 0.10 & 0.099 & 0.11 & 0.11 & 0.12 \\ \hline
  \end{tabular}
  \caption{Root mean square of radius deviations for fixed masses.}
  \label{tab:RMS}
\end{table}

For other EoSs in validation data, the corresponding $M$-$R$
curves are reconstructed well similarly to examples
discussed above.  To quantify the overall reconstruction accuracy, we
calculated the root mean square (RMS) of radius deviations using 196
validation data for several masses as shown in
Tab.~\ref{tab:RMS}.  We defined the RMS from the deviations between
not the observational data points but the genuine and reconstructed
$M$-$R$ relations (i.e.\ distances between the solid and the dashed
lines in Fig.~\ref{fig:inferred-mr}), that is,
$\delta R(M) = R^{\rm (rec)}(M)-R^{(0)}(M)$.  The RMS values in
Tab.~\ref{tab:RMS} are around $\sim 0.1\,\text{km}$ for all masses!  This
indicates that our method works surprisingly good;  remember that
data points have random fluctuations by $\Delta R\sim 0.5\,\text{km}$.
It should be noticed that, even without neutron stars around
$M=0.6$--$0.8\Msun$ in our setup, the RMS of the
corresponding radii are still reconstructed within the accuracy of the
order $\sim 0.1\,\text{km}$.
\vspace{0.3em}


Finally, let us comment on the relation to Bayesian analysis
using symbolic notations.  In our analysis we parametrized the EoS by
$\boldsymbol{\theta} := \{c_{s,i}^2\}$, which spans parameter space
$\Theta$, and generated EoSs by a probability distribution
$\Pr(\boldsymbol{\theta})$.  Then, we sampled
$\mathcal{D} = \{(M_i, R_i)\}$ by an observational distribution,
$\Pr(\mathcal{D}|\boldsymbol{\theta})$ for each EoS{}.  The neural
network is a function $f$ to obtain an EoS from data points, i.e.\
$f(\mathcal{D}|\boldsymbol{W})\in \Theta$, where
$\boldsymbol{W}$ represents the fitting parameters.  The training is
actually a process to minimize the following loss function:
\begin{equation}
  \langle \ell[f] \rangle
    = \int d\boldsymbol{\theta}d\mathcal{D}
      \Pr(\boldsymbol{\theta}) \Pr(\mathcal{D}|\boldsymbol{\theta})
      \ell(\boldsymbol{\theta}, f(\mathcal{D})).
  \label{eq:loss-in-general}
\end{equation}

Here, let us translate Bayesian analysis into the above language.
In Bayesian analysis a prior distribution of the EoS is assumed to be
$\Pr(\boldsymbol{\theta})$.  The
posterior EoS distribution is obtained by Bayesian updating;
$\Pr(\boldsymbol{\theta}|\mathcal{D}) \propto
\Pr(\boldsymbol{\theta})\Pr(\mathcal{D}|\boldsymbol{\theta})$.
To determine the most likely EoS, we can use the MAP (maximum a
posteriori) estimator,
\begin{equation}
  f_{\rm MAP}(\mathcal{D}) = \mathop{\rm arg\,max}_{\boldsymbol{\theta}}
  [\Pr(\boldsymbol{\theta})\Pr(\mathcal{D}|\boldsymbol{\theta})]\,.
\end{equation}
This can be interpreted as an approximation of $f$ that
minimizes Eq.~\eqref{eq:loss-in-general}.  This means that machine
learning encompasses Bayesian analysis as a particular limit.  Hence,
an advantage of machine learning over Bayesian analysis lies in the
direct design of the loss function or optimization target, suited for
problems under consideration.  We emphasize the generality of our
method which can be applied, with a little effort, to any
underdetermined problems;  an efficient procedure to find the most
likely solution optimized with insufficient information and limited
precision.
\vspace{0.3em}

In this work we parametrized the EoSs with five-segment piecewise
polytrope, and assumed a uniform distribution of sound velocity in
each segment to generate the training and the validation data.
We trained five-layered neural network to obtain successful results.
Important future works include systematic investigations of
performance and training costs which depend on the EoS
parametrization, training and validation data distributions, and the
neural network design.  The currently formulated method is ideal for
forthcoming neutron star observations, but for the moment the
available data spread over $M$-$R$ plane with some probability
distribution.  We are making progress to adapt our method to deal with
such data, which will be reported elsewhere.

\begin{acknowledgments}
\acknowledgments
K.~F.\ thanks Andrew~Steiner for useful discussions.
K.~F.\ and K.~M.\ are partially supported by JSPS KAKENHI Grant
No.\ 15H03652 and 15K13479.
\end{acknowledgments}

\bibliographystyle{apsrev4-1}
\bibliography{deeplearning}

\begin{thebibliography}{35}%
\makeatletter
\providecommand \@ifxundefined [1]{%
 \@ifx{#1\undefined}
}%
\providecommand \@ifnum [1]{%
 \ifnum #1\expandafter \@firstoftwo
 \else \expandafter \@secondoftwo
 \fi
}%
\providecommand \@ifx [1]{%
 \ifx #1\expandafter \@firstoftwo
 \else \expandafter \@secondoftwo
 \fi
}%
\providecommand \natexlab [1]{#1}%
\providecommand \enquote  [1]{``#1''}%
\providecommand \bibnamefont  [1]{#1}%
\providecommand \bibfnamefont [1]{#1}%
\providecommand \citenamefont [1]{#1}%
\providecommand \href@noop [0]{\@secondoftwo}%
\providecommand \href [0]{\begingroup \@sanitize@url \@href}%
\providecommand \@href[1]{\@@startlink{#1}\@@href}%
\providecommand \@@href[1]{\endgroup#1\@@endlink}%
\providecommand \@sanitize@url [0]{\catcode `\\12\catcode `\$12\catcode
  `\&12\catcode `\#12\catcode `\^12\catcode `\_12\catcode `\%12\relax}%
\providecommand \@@startlink[1]{}%
\providecommand \@@endlink[0]{}%
\providecommand \url  [0]{\begingroup\@sanitize@url \@url }%
\providecommand \@url [1]{\endgroup\@href {#1}{\urlprefix }}%
\providecommand \urlprefix  [0]{URL }%
\providecommand \Eprint [0]{\href }%
\providecommand \doibase [0]{http://dx.doi.org/}%
\providecommand \selectlanguage [0]{\@gobble}%
\providecommand \bibinfo  [0]{\@secondoftwo}%
\providecommand \bibfield  [0]{\@secondoftwo}%
\providecommand \translation [1]{[#1]}%
\providecommand \BibitemOpen [0]{}%
\providecommand \bibitemStop [0]{}%
\providecommand \bibitemNoStop [0]{.\EOS\space}%
\providecommand \EOS [0]{\spacefactor3000\relax}%
\providecommand \BibitemShut  [1]{\csname bibitem#1\endcsname}%
\let\auto@bib@innerbib\@empty
\bibitem [{\citenamefont {Aarts}(2016)}]{Aarts:2015tyj}%
  \BibitemOpen
  \bibfield  {author} {\bibinfo {author} {\bibfnamefont {G.}~\bibnamefont
  {Aarts}},\ }\bibfield  {booktitle} {\emph {\bibinfo {booktitle}
  {{Proceedings, 13th International Workshop on Hadron Physics: Angra dos Reis,
  Rio de Janeiro, Brazil, March 22-27, 2015}}},\ }\href {\doibase
  10.1088/1742-6596/706/2/022004} {\bibfield  {journal} {\bibinfo  {journal}
  {J. Phys. Conf. Ser.}\ }\textbf {\bibinfo {volume} {706}},\ \bibinfo {pages}
  {022004} (\bibinfo {year} {2016})},\ \Eprint
  {http://arxiv.org/abs/1512.05145} {arXiv:1512.05145 [hep-lat]} \BibitemShut
  {NoStop}%
\bibitem [{\citenamefont {Baym}\ \emph {et~al.}(2017)\citenamefont {Baym},
  \citenamefont {Hatsuda}, \citenamefont {Kojo}, \citenamefont {Powell},
  \citenamefont {Song},\ and\ \citenamefont {Takatsuka}}]{Baym:2017whm}%
  \BibitemOpen
  \bibfield  {author} {\bibinfo {author} {\bibfnamefont {G.}~\bibnamefont
  {Baym}}, \bibinfo {author} {\bibfnamefont {T.}~\bibnamefont {Hatsuda}},
  \bibinfo {author} {\bibfnamefont {T.}~\bibnamefont {Kojo}}, \bibinfo {author}
  {\bibfnamefont {P.~D.}\ \bibnamefont {Powell}}, \bibinfo {author}
  {\bibfnamefont {Y.}~\bibnamefont {Song}}, \ and\ \bibinfo {author}
  {\bibfnamefont {T.}~\bibnamefont {Takatsuka}},\ }\href@noop {} {\  (\bibinfo
  {year} {2017})},\ \Eprint {http://arxiv.org/abs/1707.04966} {arXiv:1707.04966
  [astro-ph.HE]} \BibitemShut {NoStop}%
\bibitem [{\citenamefont {Demorest}\ \emph {et~al.}(2010)\citenamefont
  {Demorest}, \citenamefont {Pennucci}, \citenamefont {Ransom}, \citenamefont
  {Roberts},\ and\ \citenamefont {Hessels}}]{Demorest:2010bx}%
  \BibitemOpen
  \bibfield  {author} {\bibinfo {author} {\bibfnamefont {P.}~\bibnamefont
  {Demorest}}, \bibinfo {author} {\bibfnamefont {T.}~\bibnamefont {Pennucci}},
  \bibinfo {author} {\bibfnamefont {S.}~\bibnamefont {Ransom}}, \bibinfo
  {author} {\bibfnamefont {M.}~\bibnamefont {Roberts}}, \ and\ \bibinfo
  {author} {\bibfnamefont {J.}~\bibnamefont {Hessels}},\ }\href {\doibase
  10.1038/nature09466} {\bibfield  {journal} {\bibinfo  {journal} {Nature}\
  }\textbf {\bibinfo {volume} {467}},\ \bibinfo {pages} {1081} (\bibinfo {year}
  {2010})},\ \Eprint {http://arxiv.org/abs/1010.5788} {arXiv:1010.5788
  [astro-ph.HE]} \BibitemShut {NoStop}%
\bibitem [{\citenamefont {Fonseca}\ \emph {et~al.}(2016)\citenamefont {Fonseca}
  \emph {et~al.}}]{Fonseca:2016tux}%
  \BibitemOpen
  \bibfield  {author} {\bibinfo {author} {\bibfnamefont {E.}~\bibnamefont
  {Fonseca}} \emph {et~al.},\ }\href {\doibase 10.3847/0004-637X/832/2/167}
  {\bibfield  {journal} {\bibinfo  {journal} {Astrophys. J.}\ }\textbf
  {\bibinfo {volume} {832}},\ \bibinfo {pages} {167} (\bibinfo {year}
  {2016})},\ \Eprint {http://arxiv.org/abs/1603.00545} {arXiv:1603.00545
  [astro-ph.HE]} \BibitemShut {NoStop}%
\bibitem [{\citenamefont {Antoniadis}\ \emph {et~al.}(2013)\citenamefont
  {Antoniadis} \emph {et~al.}}]{Antoniadis:2013pzd}%
  \BibitemOpen
  \bibfield  {author} {\bibinfo {author} {\bibfnamefont {J.}~\bibnamefont
  {Antoniadis}} \emph {et~al.},\ }\href {\doibase 10.1126/science.1233232}
  {\bibfield  {journal} {\bibinfo  {journal} {Science}\ }\textbf {\bibinfo
  {volume} {340}},\ \bibinfo {pages} {6131} (\bibinfo {year} {2013})},\ \Eprint
  {http://arxiv.org/abs/1304.6875} {arXiv:1304.6875 [astro-ph.HE]} \BibitemShut
  {NoStop}%
\bibitem [{\citenamefont {Fukushima}\ and\ \citenamefont
  {Hatsuda}(2011)}]{Fukushima:2010bq}%
  \BibitemOpen
  \bibfield  {author} {\bibinfo {author} {\bibfnamefont {K.}~\bibnamefont
  {Fukushima}}\ and\ \bibinfo {author} {\bibfnamefont {T.}~\bibnamefont
  {Hatsuda}},\ }\href {\doibase 10.1088/0034-4885/74/1/014001} {\bibfield
  {journal} {\bibinfo  {journal} {Rept. Prog. Phys.}\ }\textbf {\bibinfo
  {volume} {74}},\ \bibinfo {pages} {014001} (\bibinfo {year} {2011})},\
  \Eprint {http://arxiv.org/abs/1005.4814} {arXiv:1005.4814 [hep-ph]}
  \BibitemShut {NoStop}%
\bibitem [{\citenamefont {Schmidt}\ and\ \citenamefont
  {Sharma}(2017)}]{Schmidt:2017bjt}%
  \BibitemOpen
  \bibfield  {author} {\bibinfo {author} {\bibfnamefont {C.}~\bibnamefont
  {Schmidt}}\ and\ \bibinfo {author} {\bibfnamefont {S.}~\bibnamefont
  {Sharma}},\ }\href {\doibase 10.1088/1361-6471/aa824a} {\bibfield  {journal}
  {\bibinfo  {journal} {J. Phys.}\ }\textbf {\bibinfo {volume} {G44}},\
  \bibinfo {pages} {104002} (\bibinfo {year} {2017})},\ \Eprint
  {http://arxiv.org/abs/1701.04707} {arXiv:1701.04707 [hep-lat]} \BibitemShut
  {NoStop}%
\bibitem [{\citenamefont {Masuda}\ \emph {et~al.}(2013)\citenamefont {Masuda},
  \citenamefont {Hatsuda},\ and\ \citenamefont {Takatsuka}}]{Masuda:2012kf}%
  \BibitemOpen
  \bibfield  {author} {\bibinfo {author} {\bibfnamefont {K.}~\bibnamefont
  {Masuda}}, \bibinfo {author} {\bibfnamefont {T.}~\bibnamefont {Hatsuda}}, \
  and\ \bibinfo {author} {\bibfnamefont {T.}~\bibnamefont {Takatsuka}},\ }\href
  {\doibase 10.1088/0004-637X/764/1/12} {\bibfield  {journal} {\bibinfo
  {journal} {Astrophys. J.}\ }\textbf {\bibinfo {volume} {764}},\ \bibinfo
  {pages} {12} (\bibinfo {year} {2013})},\ \Eprint
  {http://arxiv.org/abs/1205.3621} {arXiv:1205.3621 [nucl-th]} \BibitemShut
  {NoStop}%
\bibitem [{\citenamefont {Alvarez-Castillo}\ \emph {et~al.}(2014)\citenamefont
  {Alvarez-Castillo}, \citenamefont {Benic}, \citenamefont {Blaschke},\ and\
  \citenamefont {Łastowiecki}}]{Alvarez-Castillo:2013spa}%
  \BibitemOpen
  \bibfield  {author} {\bibinfo {author} {\bibfnamefont {D.~E.}\ \bibnamefont
  {Alvarez-Castillo}}, \bibinfo {author} {\bibfnamefont {S.}~\bibnamefont
  {Benic}}, \bibinfo {author} {\bibfnamefont {D.}~\bibnamefont {Blaschke}}, \
  and\ \bibinfo {author} {\bibfnamefont {R.}~\bibnamefont {Łastowiecki}},\
  }\bibfield  {booktitle} {\emph {\bibinfo {booktitle} {{Proceedings, 31st Max
  Born Symposium and HIC for FAIR Workshop : Three Days of critical behaviour
  in hot and dense QCD: Wroclaw, Poland, June 14-16, 2013}}},\ }\href {\doibase
  10.5506/APhysPolBSupp.7.203} {\bibfield  {journal} {\bibinfo  {journal} {Acta
  Phys. Polon. Supp.}\ }\textbf {\bibinfo {volume} {7}},\ \bibinfo {pages}
  {203} (\bibinfo {year} {2014})},\ \Eprint {http://arxiv.org/abs/1311.5112}
  {arXiv:1311.5112 [nucl-th]} \BibitemShut {NoStop}%
\bibitem [{\citenamefont {Fukushima}\ and\ \citenamefont
  {Kojo}(2016)}]{Fukushima:2015bda}%
  \BibitemOpen
  \bibfield  {author} {\bibinfo {author} {\bibfnamefont {K.}~\bibnamefont
  {Fukushima}}\ and\ \bibinfo {author} {\bibfnamefont {T.}~\bibnamefont
  {Kojo}},\ }\href {\doibase 10.3847/0004-637X/817/2/180} {\bibfield  {journal}
  {\bibinfo  {journal} {Astrophys. J.}\ }\textbf {\bibinfo {volume} {817}},\
  \bibinfo {pages} {180} (\bibinfo {year} {2016})},\ \Eprint
  {http://arxiv.org/abs/1509.00356} {arXiv:1509.00356 [nucl-th]} \BibitemShut
  {NoStop}%
\bibitem [{\citenamefont {{Sch\"afer}}\ and\ \citenamefont
  {Wilczek}(1999)}]{Schafer:1998ef}%
  \BibitemOpen
  \bibfield  {author} {\bibinfo {author} {\bibfnamefont {T.}~\bibnamefont
  {{Sch\"afer}}}\ and\ \bibinfo {author} {\bibfnamefont {F.}~\bibnamefont
  {Wilczek}},\ }\href {\doibase 10.1103/PhysRevLett.82.3956} {\bibfield
  {journal} {\bibinfo  {journal} {Phys. Rev. Lett.}\ }\textbf {\bibinfo
  {volume} {82}},\ \bibinfo {pages} {3956} (\bibinfo {year} {1999})},\ \Eprint
  {http://arxiv.org/abs/hep-ph/9811473} {arXiv:hep-ph/9811473 [hep-ph]}
  \BibitemShut {NoStop}%
\bibitem [{\citenamefont {Alford}\ \emph {et~al.}(1999)\citenamefont {Alford},
  \citenamefont {Berges},\ and\ \citenamefont {Rajagopal}}]{Alford:1999pa}%
  \BibitemOpen
  \bibfield  {author} {\bibinfo {author} {\bibfnamefont {M.~G.}\ \bibnamefont
  {Alford}}, \bibinfo {author} {\bibfnamefont {J.}~\bibnamefont {Berges}}, \
  and\ \bibinfo {author} {\bibfnamefont {K.}~\bibnamefont {Rajagopal}},\ }\href
  {\doibase 10.1016/S0550-3213(99)00410-1} {\bibfield  {journal} {\bibinfo
  {journal} {Nucl. Phys.}\ }\textbf {\bibinfo {volume} {B558}},\ \bibinfo
  {pages} {219} (\bibinfo {year} {1999})},\ \Eprint
  {http://arxiv.org/abs/hep-ph/9903502} {arXiv:hep-ph/9903502 [hep-ph]}
  \BibitemShut {NoStop}%
\bibitem [{\citenamefont {Fukushima}(2004)}]{Fukushima:2004bj}%
  \BibitemOpen
  \bibfield  {author} {\bibinfo {author} {\bibfnamefont {K.}~\bibnamefont
  {Fukushima}},\ }\href {\doibase 10.1103/PhysRevD.70.094014} {\bibfield
  {journal} {\bibinfo  {journal} {Phys. Rev.}\ }\textbf {\bibinfo {volume}
  {D70}},\ \bibinfo {pages} {094014} (\bibinfo {year} {2004})},\ \Eprint
  {http://arxiv.org/abs/hep-ph/0403091} {arXiv:hep-ph/0403091 [hep-ph]}
  \BibitemShut {NoStop}%
\bibitem [{\citenamefont {Hatsuda}\ \emph {et~al.}(2006)\citenamefont
  {Hatsuda}, \citenamefont {Tachibana}, \citenamefont {Yamamoto},\ and\
  \citenamefont {Baym}}]{Hatsuda:2006ps}%
  \BibitemOpen
  \bibfield  {author} {\bibinfo {author} {\bibfnamefont {T.}~\bibnamefont
  {Hatsuda}}, \bibinfo {author} {\bibfnamefont {M.}~\bibnamefont {Tachibana}},
  \bibinfo {author} {\bibfnamefont {N.}~\bibnamefont {Yamamoto}}, \ and\
  \bibinfo {author} {\bibfnamefont {G.}~\bibnamefont {Baym}},\ }\href {\doibase
  10.1103/PhysRevLett.97.122001} {\bibfield  {journal} {\bibinfo  {journal}
  {Phys. Rev. Lett.}\ }\textbf {\bibinfo {volume} {97}},\ \bibinfo {pages}
  {122001} (\bibinfo {year} {2006})},\ \Eprint
  {http://arxiv.org/abs/hep-ph/0605018} {arXiv:hep-ph/0605018 [hep-ph]}
  \BibitemShut {NoStop}%
\bibitem [{\citenamefont {{\"Ozel}}\ and\ \citenamefont
  {Freire}(2016)}]{Ozel:2016oaf}%
  \BibitemOpen
  \bibfield  {author} {\bibinfo {author} {\bibfnamefont {F.}~\bibnamefont
  {{\"Ozel}}}\ and\ \bibinfo {author} {\bibfnamefont {P.}~\bibnamefont
  {Freire}},\ }\href {\doibase 10.1146/annurev-astro-081915-023322} {\bibfield
  {journal} {\bibinfo  {journal} {Ann. Rev. Astron. Astrophys.}\ }\textbf
  {\bibinfo {volume} {54}},\ \bibinfo {pages} {401} (\bibinfo {year} {2016})},\
  \Eprint {http://arxiv.org/abs/1603.02698} {arXiv:1603.02698 [astro-ph.HE]}
  \BibitemShut {NoStop}%
\bibitem [{\citenamefont {Tolman}(1939)}]{Tolman:1939jz}%
  \BibitemOpen
  \bibfield  {author} {\bibinfo {author} {\bibfnamefont {R.~C.}\ \bibnamefont
  {Tolman}},\ }\href {\doibase 10.1103/PhysRev.55.364} {\bibfield  {journal}
  {\bibinfo  {journal} {Phys. Rev.}\ }\textbf {\bibinfo {volume} {55}},\
  \bibinfo {pages} {364} (\bibinfo {year} {1939})}\BibitemShut {NoStop}%
\bibitem [{\citenamefont {Oppenheimer}\ and\ \citenamefont
  {Volkoff}(1939)}]{Oppenheimer:1939ne}%
  \BibitemOpen
  \bibfield  {author} {\bibinfo {author} {\bibfnamefont {J.~R.}\ \bibnamefont
  {Oppenheimer}}\ and\ \bibinfo {author} {\bibfnamefont {G.~M.}\ \bibnamefont
  {Volkoff}},\ }\href {\doibase 10.1103/PhysRev.55.374} {\bibfield  {journal}
  {\bibinfo  {journal} {Phys. Rev.}\ }\textbf {\bibinfo {volume} {55}},\
  \bibinfo {pages} {374} (\bibinfo {year} {1939})}\BibitemShut {NoStop}%
\bibitem [{\citenamefont {Sotani}\ and\ \citenamefont
  {Tatsumi}(2015)}]{Sotani:2014rva}%
  \BibitemOpen
  \bibfield  {author} {\bibinfo {author} {\bibfnamefont {H.}~\bibnamefont
  {Sotani}}\ and\ \bibinfo {author} {\bibfnamefont {T.}~\bibnamefont
  {Tatsumi}},\ }\href {\doibase 10.1093/mnras/stu2677} {\bibfield  {journal}
  {\bibinfo  {journal} {Mon. Not. Roy. Astron. Soc.}\ }\textbf {\bibinfo
  {volume} {447}},\ \bibinfo {pages} {3155} (\bibinfo {year} {2015})},\ \Eprint
  {http://arxiv.org/abs/1412.4610} {arXiv:1412.4610 [astro-ph.HE]} \BibitemShut
  {NoStop}%
\bibitem [{\citenamefont {{Lindblom}}(1992)}]{Lindblom:1992}%
  \BibitemOpen
  \bibfield  {author} {\bibinfo {author} {\bibfnamefont {L.}~\bibnamefont
  {{Lindblom}}},\ }\href {\doibase 10.1086/171882} {\bibfield  {journal}
  {\bibinfo  {journal} {\apj}\ }\textbf {\bibinfo {volume} {398}},\ \bibinfo
  {pages} {569} (\bibinfo {year} {1992})}\BibitemShut {NoStop}%
\bibitem [{\citenamefont {Schertler}\ \emph {et~al.}(2000)\citenamefont
  {Schertler}, \citenamefont {Greiner}, \citenamefont {Schaffner-Bielich},\
  and\ \citenamefont {Thoma}}]{Schertler:2000xq}%
  \BibitemOpen
  \bibfield  {author} {\bibinfo {author} {\bibfnamefont {K.}~\bibnamefont
  {Schertler}}, \bibinfo {author} {\bibfnamefont {C.}~\bibnamefont {Greiner}},
  \bibinfo {author} {\bibfnamefont {J.}~\bibnamefont {Schaffner-Bielich}}, \
  and\ \bibinfo {author} {\bibfnamefont {M.~H.}\ \bibnamefont {Thoma}},\ }\href
  {\doibase 10.1016/S0375-9474(00)00305-5} {\bibfield  {journal} {\bibinfo
  {journal} {Nucl. Phys.}\ }\textbf {\bibinfo {volume} {A677}},\ \bibinfo
  {pages} {463} (\bibinfo {year} {2000})},\ \Eprint
  {http://arxiv.org/abs/astro-ph/0001467} {arXiv:astro-ph/0001467 [astro-ph]}
  \BibitemShut {NoStop}%
\bibitem [{\citenamefont {Alford}\ and\ \citenamefont
  {Sedrakian}(2017)}]{Alford:2017qgh}%
  \BibitemOpen
  \bibfield  {author} {\bibinfo {author} {\bibfnamefont {M.~G.}\ \bibnamefont
  {Alford}}\ and\ \bibinfo {author} {\bibfnamefont {A.}~\bibnamefont
  {Sedrakian}},\ }\href {\doibase 10.1103/PhysRevLett.119.161104} {\bibfield
  {journal} {\bibinfo  {journal} {Phys. Rev. Lett.}\ }\textbf {\bibinfo
  {volume} {119}},\ \bibinfo {pages} {161104} (\bibinfo {year} {2017})},\
  \Eprint {http://arxiv.org/abs/1706.01592} {arXiv:1706.01592 [astro-ph.HE]}
  \BibitemShut {NoStop}%
\bibitem [{\citenamefont {Lindblom}(2010)}]{Lindblom:2010bb}%
  \BibitemOpen
  \bibfield  {author} {\bibinfo {author} {\bibfnamefont {L.}~\bibnamefont
  {Lindblom}},\ }\href {\doibase 10.1103/PhysRevD.82.103011} {\bibfield
  {journal} {\bibinfo  {journal} {Phys. Rev.}\ }\textbf {\bibinfo {volume}
  {D82}},\ \bibinfo {pages} {103011} (\bibinfo {year} {2010})},\ \Eprint
  {http://arxiv.org/abs/1009.0738} {arXiv:1009.0738 [astro-ph.HE]} \BibitemShut
  {NoStop}%
\bibitem [{\citenamefont {Lindblom}\ and\ \citenamefont
  {Indik}(2014)}]{Lindblom:2013kra}%
  \BibitemOpen
  \bibfield  {author} {\bibinfo {author} {\bibfnamefont {L.}~\bibnamefont
  {Lindblom}}\ and\ \bibinfo {author} {\bibfnamefont {N.~M.}\ \bibnamefont
  {Indik}},\ }\href {\doibase 10.1103/PhysRevD.89.064003,
  10.1103/PhysRevD.93.129903} {\bibfield  {journal} {\bibinfo  {journal} {Phys.
  Rev.}\ }\textbf {\bibinfo {volume} {D89}},\ \bibinfo {pages} {064003}
  (\bibinfo {year} {2014})},\ \bibinfo {note} {[Erratum: Phys.
  Rev.D93,no.12,129903(2016)]},\ \Eprint {http://arxiv.org/abs/1310.0803}
  {arXiv:1310.0803 [astro-ph.HE]} \BibitemShut {NoStop}%
\bibitem [{\citenamefont {Read}\ \emph {et~al.}(2009)\citenamefont {Read},
  \citenamefont {Lackey}, \citenamefont {Owen},\ and\ \citenamefont
  {Friedman}}]{Read:2008iy}%
  \BibitemOpen
  \bibfield  {author} {\bibinfo {author} {\bibfnamefont {J.~S.}\ \bibnamefont
  {Read}}, \bibinfo {author} {\bibfnamefont {B.~D.}\ \bibnamefont {Lackey}},
  \bibinfo {author} {\bibfnamefont {B.~J.}\ \bibnamefont {Owen}}, \ and\
  \bibinfo {author} {\bibfnamefont {J.~L.}\ \bibnamefont {Friedman}},\ }\href
  {\doibase 10.1103/PhysRevD.79.124032} {\bibfield  {journal} {\bibinfo
  {journal} {Phys. Rev.}\ }\textbf {\bibinfo {volume} {D79}},\ \bibinfo {pages}
  {124032} (\bibinfo {year} {2009})},\ \Eprint {http://arxiv.org/abs/0812.2163}
  {arXiv:0812.2163 [astro-ph]} \BibitemShut {NoStop}%
\bibitem [{\citenamefont {{\"Ozel}}\ \emph {et~al.}(2010)\citenamefont
  {{\"Ozel}}, \citenamefont {Baym},\ and\ \citenamefont {Guver}}]{Ozel:2010fw}%
  \BibitemOpen
  \bibfield  {author} {\bibinfo {author} {\bibfnamefont {F.}~\bibnamefont
  {{\"Ozel}}}, \bibinfo {author} {\bibfnamefont {G.}~\bibnamefont {Baym}}, \
  and\ \bibinfo {author} {\bibfnamefont {T.}~\bibnamefont {Guver}},\ }\href
  {\doibase 10.1103/PhysRevD.82.101301} {\bibfield  {journal} {\bibinfo
  {journal} {Phys. Rev.}\ }\textbf {\bibinfo {volume} {D82}},\ \bibinfo {pages}
  {101301} (\bibinfo {year} {2010})},\ \Eprint {http://arxiv.org/abs/1002.3153}
  {arXiv:1002.3153 [astro-ph.HE]} \BibitemShut {NoStop}%
\bibitem [{\citenamefont {{Raithel, Carolyn A. and {\"Ozel}, Feryal and
  Psaltis, Dimitrios}}(2017)}]{Raithel:2017ity}%
  \BibitemOpen
  \bibfield  {author} {\bibinfo {author} {\bibnamefont {{Raithel, Carolyn A.
  and {\"Ozel}, Feryal and Psaltis, Dimitrios}}},\ }\href {\doibase
  10.3847/1538-4357/aa7a5a} {\bibfield  {journal} {\bibinfo  {journal}
  {Astrophys. J.}\ }\textbf {\bibinfo {volume} {844}},\ \bibinfo {pages} {156}
  (\bibinfo {year} {2017})},\ \Eprint {http://arxiv.org/abs/1704.00737}
  {arXiv:1704.00737 [astro-ph.HE]} \BibitemShut {NoStop}%
\bibitem [{\citenamefont {Steiner}\ \emph {et~al.}(2010)\citenamefont
  {Steiner}, \citenamefont {Lattimer},\ and\ \citenamefont
  {Brown}}]{Steiner:2010fz}%
  \BibitemOpen
  \bibfield  {author} {\bibinfo {author} {\bibfnamefont {A.~W.}\ \bibnamefont
  {Steiner}}, \bibinfo {author} {\bibfnamefont {J.~M.}\ \bibnamefont
  {Lattimer}}, \ and\ \bibinfo {author} {\bibfnamefont {E.~F.}\ \bibnamefont
  {Brown}},\ }\href {\doibase 10.1088/0004-637X/722/1/33} {\bibfield  {journal}
  {\bibinfo  {journal} {Astrophys. J.}\ }\textbf {\bibinfo {volume} {722}},\
  \bibinfo {pages} {33} (\bibinfo {year} {2010})},\ \Eprint
  {http://arxiv.org/abs/1005.0811} {arXiv:1005.0811 [astro-ph.HE]} \BibitemShut
  {NoStop}%
\bibitem [{\citenamefont {Alvarez-Castillo}\ \emph {et~al.}(2016)\citenamefont
  {Alvarez-Castillo}, \citenamefont {Ayriyan}, \citenamefont {Benic},
  \citenamefont {Blaschke}, \citenamefont {Grigorian},\ and\ \citenamefont
  {Typel}}]{Alvarez-Castillo:2016oln}%
  \BibitemOpen
  \bibfield  {author} {\bibinfo {author} {\bibfnamefont {D.}~\bibnamefont
  {Alvarez-Castillo}}, \bibinfo {author} {\bibfnamefont {A.}~\bibnamefont
  {Ayriyan}}, \bibinfo {author} {\bibfnamefont {S.}~\bibnamefont {Benic}},
  \bibinfo {author} {\bibfnamefont {D.}~\bibnamefont {Blaschke}}, \bibinfo
  {author} {\bibfnamefont {H.}~\bibnamefont {Grigorian}}, \ and\ \bibinfo
  {author} {\bibfnamefont {S.}~\bibnamefont {Typel}},\ }\href {\doibase
  10.1140/epja/i2016-16069-2} {\bibfield  {journal} {\bibinfo  {journal} {Eur.
  Phys. J.}\ }\textbf {\bibinfo {volume} {A52}},\ \bibinfo {pages} {69}
  (\bibinfo {year} {2016})},\ \Eprint {http://arxiv.org/abs/1603.03457}
  {arXiv:1603.03457 [nucl-th]} \BibitemShut {NoStop}%
\bibitem [{\citenamefont {Pang}\ \emph {et~al.}(2016)\citenamefont {Pang},
  \citenamefont {Zhou}, \citenamefont {Su}, \citenamefont {Petersen},
  \citenamefont {St{\"o}cker},\ and\ \citenamefont {Wang}}]{Pang:2016vdc}%
  \BibitemOpen
  \bibfield  {author} {\bibinfo {author} {\bibfnamefont {L.-G.}\ \bibnamefont
  {Pang}}, \bibinfo {author} {\bibfnamefont {K.}~\bibnamefont {Zhou}}, \bibinfo
  {author} {\bibfnamefont {N.}~\bibnamefont {Su}}, \bibinfo {author}
  {\bibfnamefont {H.}~\bibnamefont {Petersen}}, \bibinfo {author}
  {\bibfnamefont {H.}~\bibnamefont {St{\"o}cker}}, \ and\ \bibinfo {author}
  {\bibfnamefont {X.-N.}\ \bibnamefont {Wang}},\ }\href@noop {} {\  (\bibinfo
  {year} {2016})},\ \Eprint {http://arxiv.org/abs/1612.04262} {arXiv:1612.04262
  [hep-ph]} \BibitemShut {NoStop}%
\bibitem [{\citenamefont {Niu}\ and\ \citenamefont
  {Liang}(2018)}]{Niu:2018csp}%
  \BibitemOpen
  \bibfield  {author} {\bibinfo {author} {\bibfnamefont {Z.~M.}\ \bibnamefont
  {Niu}}\ and\ \bibinfo {author} {\bibfnamefont {H.~Z.}\ \bibnamefont
  {Liang}},\ }\href {\doibase 10.1016/j.physletb.2018.01.002} {\bibfield
  {journal} {\bibinfo  {journal} {Phys. Lett.}\ }\textbf {\bibinfo {volume}
  {B778}},\ \bibinfo {pages} {48} (\bibinfo {year} {2018})},\ \Eprint
  {http://arxiv.org/abs/1801.04411} {arXiv:1801.04411 [nucl-th]} \BibitemShut
  {NoStop}%
\bibitem [{\citenamefont {{Raithel, Carolyn A. and {\"Ozel}, Feryal and
  Psaltis, Dimitrios}}(2016)}]{Raithel:2016bux}%
  \BibitemOpen
  \bibfield  {author} {\bibinfo {author} {\bibnamefont {{Raithel, Carolyn A.
  and {\"Ozel}, Feryal and Psaltis, Dimitrios}}},\ }\href {\doibase
  10.3847/0004-637X/831/1/44} {\bibfield  {journal} {\bibinfo  {journal}
  {Astrophys. J.}\ }\textbf {\bibinfo {volume} {831}},\ \bibinfo {pages} {44}
  (\bibinfo {year} {2016})},\ \Eprint {http://arxiv.org/abs/1605.03591}
  {arXiv:1605.03591 [astro-ph.HE]} \BibitemShut {NoStop}%
\bibitem [{\citenamefont {Douchin}\ and\ \citenamefont
  {Haensel}(2001)}]{Douchin:2001sv}%
  \BibitemOpen
  \bibfield  {author} {\bibinfo {author} {\bibfnamefont {F.}~\bibnamefont
  {Douchin}}\ and\ \bibinfo {author} {\bibfnamefont {P.}~\bibnamefont
  {Haensel}},\ }\href {\doibase 10.1051/0004-6361:20011402} {\bibfield
  {journal} {\bibinfo  {journal} {Astron. Astrophys.}\ }\textbf {\bibinfo
  {volume} {380}},\ \bibinfo {pages} {151} (\bibinfo {year} {2001})},\ \Eprint
  {http://arxiv.org/abs/astro-ph/0111092} {arXiv:astro-ph/0111092 [astro-ph]}
  \BibitemShut {NoStop}%
\bibitem [{\citenamefont {Chollet}(2015)}]{software:Keras}%
  \BibitemOpen
  \bibfield  {author} {\bibinfo {author} {\bibfnamefont {F.}~\bibnamefont
  {Chollet}},\ }\href@noop {} {\enquote {\bibinfo {title} {Keras: Deep learning
  library for theano and tensorflow},}\ }\bibinfo {howpublished}
  {\url{https://github.com/fchollet/keras}} (\bibinfo {year}
  {2015})\BibitemShut {NoStop}%
\bibitem [{\citenamefont {Abadi}\ \emph {et~al.}(2016)\citenamefont {Abadi}
  \emph {et~al.}}]{arXiv:1605.08695}%
  \BibitemOpen
  \bibfield  {author} {\bibinfo {author} {\bibfnamefont {M.}~\bibnamefont
  {Abadi}} \emph {et~al.},\ }\href@noop {} {\  (\bibinfo {year} {2016})},\
  \Eprint {http://arxiv.org/abs/1605.08695} {arXiv:1605.08695 [cs.DC]}
  \BibitemShut {NoStop}%
\bibitem [{\citenamefont {Kingma}\ and\ \citenamefont
  {Ba}(2014)}]{DBLP:journals/corr/KingmaB14}%
  \BibitemOpen
  \bibfield  {author} {\bibinfo {author} {\bibfnamefont {D.~P.}\ \bibnamefont
  {Kingma}}\ and\ \bibinfo {author} {\bibfnamefont {J.}~\bibnamefont {Ba}},\
  }\href {http://arxiv.org/abs/1412.6980} {\bibfield  {journal} {\bibinfo
  {journal} {CoRR}\ }\textbf {\bibinfo {volume} {abs/1412.6980}} (\bibinfo
  {year} {2014})},\ \Eprint {http://arxiv.org/abs/1412.6980} {arXiv:1412.6980}
  \BibitemShut {NoStop}%
\end{thebibliography}%

\end{document}